\documentclass{mn2e}
\begin{document}

\newcommand\etal{et~al.}
\newcommand\methanol{$\rm{CH_{3}OH}$}
\newcommand\six{$5_{1}-6_{0}\ \rm{A}^{+}$}
\newcommand\twelve{$2_{0}-3_{-1}\ \rm{E}$}
\newcommand\vt{\mbox{$v_{\rm t}$}}
\newcommand\vCO{\mbox{$v_{\rm CO}$}}
\newcommand\wn{\mbox{$\rm cm^{-1}$}}
\newcommand\rate{\mbox{$\rm cm^{3} s^{-1}$}}
\newcommand\nH{\mbox{$n_{\rm H}$}}
\newcommand\scdm{\mbox{$N_{\rm M}/\Delta V$}}
\newcommand\Td{\mbox{$T_{\rm d}$}}
\newcommand\Tk{\mbox{$T_{\rm k}$}}
\newcommand\Xm{\mbox{$X_{\rm M}$}}
\newcommand\beam{\mbox{$\epsilon^{-1}$}}
\newcommand\DV{\mbox{$\Delta V$}}
\newcommand\HII{H\,{\sc ii}}
\newcommand\WHII{\mbox{$W_{\rm HII}$}}
\newcommand\cc{\mbox{$\,\rm cm^{-3}$}}
\newcommand\ccs{\mbox{$\,\rm cm^{-3} s$}}
\newcommand\kms{\mbox{${\rm km\,s}^{-1}$}}
\newcommand\Wd{\mbox{$W_{\rm d}$}}
\newcommand\taud{\mbox{$\tau_{\rm d}$}}
\newcommand\Te{\mbox{$T_{\rm e}$}}
\newcommand\fe{\mbox{$f_{\rm e}$}}
\newcommand\Tb{\mbox{$T_{\rm b}$}}

\input epsf
\epsfverbosetrue

\title[Class II methanol masers]{Models of class~II methanol masers based on improved molecular data}

\author[Cragg \etal\/]{D.M. Cragg$^1$, A.M. Sobolev$^2$, and P.D. Godfrey$^1$\\
$^1$ School of Chemistry, Building 23, Monash University, Victoria 3800,  Australia;\\ Dinah.Cragg@sci.monash.edu.au, Peter.Godfrey@sci.monash.edu.au\\
$^2$ Astronomical Observatory, Ural State University, Lenin Street 51, Ekaterinburg 620083, Russia;\\ Andrej.Sobolev@usu.ru,\\}

\maketitle


\begin{abstract}

The class~II masers of methanol are associated with the early stages of formation of high-mass stars.  Modelling of these dense, dusty environments has demonstrated that pumping by infrared radiation can account for the observed masers.  Collisions with other molecules in the ambient gas also play a significant role, but have not been well modelled in the past.  Here we examine the effects on the maser models of newly available collision rate coefficients for methanol.  The new collision data does not alter which transitions become masers in the models, but does influence their brightness and the conditions under which they switch on and off.  At gas temperatures above 100~K the effects are broadly consistent with a reduction in the overall collision cross-section.  This means, for example, that a slightly higher gas density than identified previously can account for most of the observed masers in W3(OH).  We have also examined the effects of including more excited state energy levels in the models, and find that these play a significant role only at dust temperatures above 300~K.  An updated list of class~II methanol maser candidates is presented.

\end{abstract}



\section{INTRODUCTION}

Methanol, \methanol, is an important constituent of the interstellar gas in star-forming regions.  It has a rich microwave and millimetre spectrum as a result of its asymmetry, and low-energy torsional vibrations of the methyl group against the OH frame can be excited under interstellar conditions.  Radioastronomical observations of methanol provide a useful probe of physical and chemical conditions.  This is particularly true in regions of high-mass star formation, where a large gas-phase abundance results from evaporation of solid state methanol from grain mantles.  Furthermore, the physical conditions in such environments give rise to masers, as well as thermal and quasi-thermal emission.  The interpretation of all such emissions requires reliable models of the excitation processes.

Until now, the biggest deficiency in models of methanol exciation has been the lack of reliable rate coefficients for collisions with other constituents of the gas.  Most authors have either stuck to a Local Thermodynamic Equilibrium approach (assuming a Boltzmann population distribution), or have used propensity rules to approximate the collision rate coefficients, based on a small number of double resonance experiments by Lees \& Haque (1974).  Recently, the first accurately calculated rate coefficients for methanol excitation by collisions with He and para-H$_{2}$ have become available (Pottage, Flower \& Davis 2001, 2002, 2004a, 2004b, hereafter PFD).  These give the prospect of obtaining more accurate information from observations of methanol emissions in a variety of environments.

Here we focus on the class~II masers of methanol, which are a tracer of recent high-mass star formation.  They are often closely associated with OH masers as well as infrared and uc\HII\ region continuum emission from such regions.  They pinpoint the very early development of a high-mass star, and may in fact pre-date the development of other tracers.  Their class~I maser cousins are also found in star-formation regions, but are ususally well separated from the source of excitation.  The most prevalent class~II methanol maser line is the \six\ transition at 6.668~GHz, known at some 500 sites (see catalogs of Malyshev \& Sobolev 2003; Pestalozzi, Minier \& Booth 2004).  The \twelve\ transition at 12.172~GHz has also been detected at many of these locations.  There are around 20 further class~II methanol maser transitions which have features in common with the 6.6- and 12.1-GHz masers; these are generally weaker and known in only a small number of sources.

The class~II masers can be accounted for by the model of Sobolev \& Deguchi (1994), in which infrared radiation pumps methanol molecules to the second torsionally excited state.  While the excitation is predominantly radiative, collisional quenching becomes significant at high densities.  In this paper we revisit past work with this model (Sobolev, Cragg \& Godfrey 1997a, 1997b; Cragg \etal\ 2001; Sutton \etal\ 2001; Cragg \etal\ 2004) to ascertain the impact of the new collision data.  As well, we look at the effect of including more energy levels in the calculations, and specifically more torsionally and vibrationally excited states.


\section{MASER MODELLING}

In the spectrum of an interstellar molecule, the relative intensity in different lines reflects the physical conditions under which they were produced.  At low interstellar densities the populations of the various energy levels usually do not follow a Boltzmann distribution, but are determined by a statistical equilibrium between excitation and de-excitation processes.  Molecules move from one energy state to another by radiative (absorption, spontaneous or stimulated emission of photons) or collisional means (due to interactions with other constituents of the gas, principally H$_{2}$ and He).  Methanol masers arise when the competition between these processes produces population inversions between the irregularly spaced energy levels, and when there is enough methanol present that amplification can develop along particular lines of sight.  In practice, when modelling these processes, the self-consistent calculation of level populations from the transition rates can only be done for a finite number of energy levels, and if too few are inlcuded the population figures become distorted.  The models rely on the avalability of high-quality molecular data for the energies of the various accessible quantum states, for the Einstein coefficients or line strengths which determine the rates of radiative transitions, and for the collisional rate coefficients.  The currently available molecular data for methanol is described in the next two subsections, followed by a description of the class~II maser model.


\subsection{Energy levels of methanol}


\begin{figure*}
\centerline{\epsfxsize=18cm\epsfbox{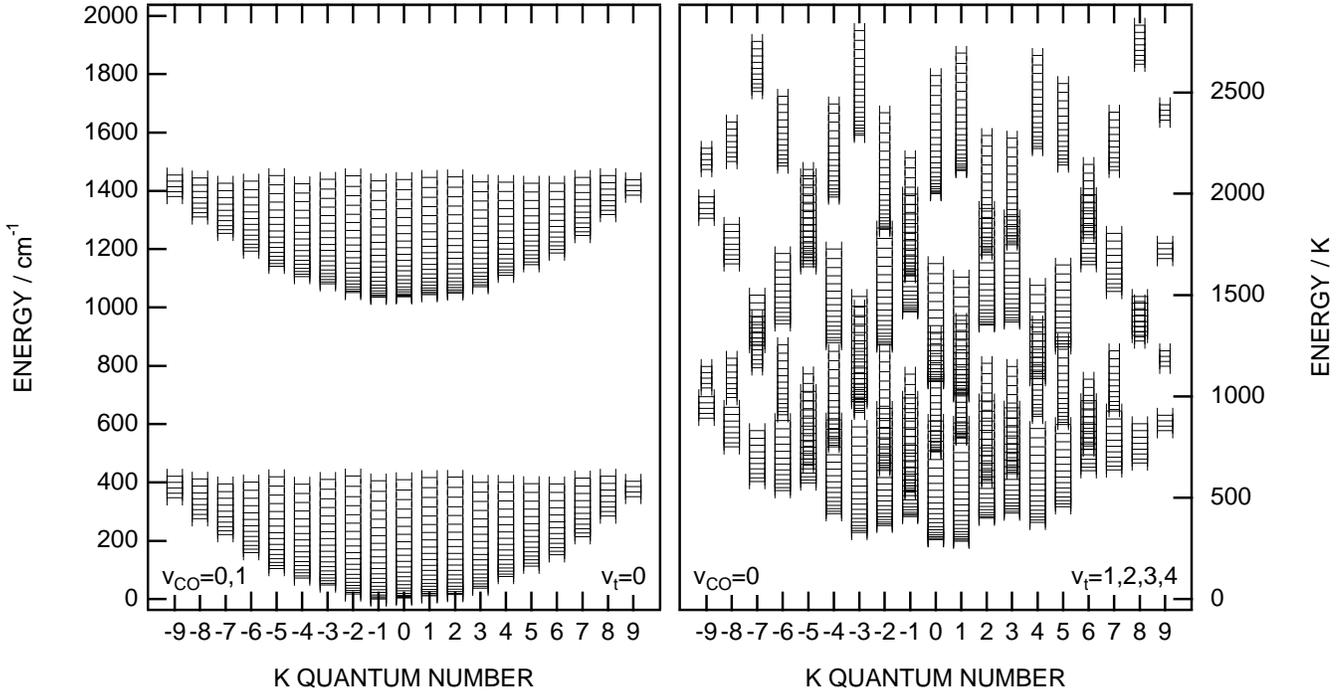}}
\caption{Energy levels used in the largest maser model calculations for the E symmetry species of methanol.  Left panel shows levels of the ground vibrational state,  including rotational states up to an energy threshold with maximum J= 22, plus corresponding levels of the CO-stretch vibration.  Right panel shows corresponding levels of the first 4 torsionally excited states.  For clarity, $\vt=2$ and 4  are shown as heavier and narrower ladders compared with $\vt=1$ and 3.  All levels are on the same scale and referred to the same ground state, with energy in units of cm$^{-1}$ on left axis and K on right axis.}
\label{fig:levels}
\end{figure*}


Methanol molecules exist in two symmetry states, denoted A and E, which differ in the alignment of the nuclear spins.  As the normal radiative and collisional processes do not interconvert these symmetry states, they are treated independently.  The A and E species are usually assumed to be equally abundant, although an excess of the A species follows from its lower ground-state energy if the molecules formed under very cold conditions (e.g. [A]/[E]=1.44 if formed at 10~K).  This is a possibility in star-forming regions where the bulk of the gas-phase methanol is believed to have evaporated from grain surfaces, where it formed or was deposited during an earlier evolutionary stage.

The methanol eigenstates are labelled by the overall rotational angular momentum quantum number J.  All radiative transitions are subject to the selection rule $\Delta$J=0,$\pm1$.  Conventionally the states are also labelled by K, representing the projection of J onto the axis of symmetry, although the asymmetry and the internal rotation mean that K is not a strict quantum number (as it would be for a rigid symmetric rotor).  The A symmetry species has close pairing of levels, which are labelled by a $\pm$ symmetry label (related to the parity quantum number) and nonnegative K, while the E species levels are instead labelled by a signed K (see Lees \& Baker 1968).  The strongest radiative transitions have $\Delta$K=0,$\pm1$.  Over 700 methanol rotational lines have been identified in interstellar surveys (Lovas 2004), with quantum numbers ranging from J=0 to J=26.

The lowest energy vibrational motion in methanol is the torsion or hindered internal rotation of the methyl group against the OH frame (\vt=1), which is excited at energy approximately 200~\wn\ or 300~K above the ground state.  Over 200 lines from the first and second torsionally excited states (\vt=1,2) are known in warm interstellar sources such as Orion (Lovas 2004).  This suggests either that the gas is warm enough that such high levels are populated by thermal means, or that molecules have absorbed infrared radiation at around 50 $\mu$m to populate the torsionally excited states.  Radiative transitions between different torsional states with $\Delta$K=$\pm1$ are strongly allowed due to the large degree of torsion-rotation interaction in methanol.  The sequence of higher torsional states (\vt=1,2,3,4 etc) each has its associated manifold of rotational levels (see Fig~\ref{fig:levels}).  The next vibrational mode is the CO stretch (\vCO=1), which appears at energy approximately 1000~\wn\ above the ground state, closely followed by the CH$_{3}$ rocking mode.

The infrared signatures of star-forming regions associated with class~II methanol masers typically indicate dust temperatures 100--200~K (De Buizer, Pina \& Telesco 2000), and it seems likely that the gas temperatures are of the same order.  Even without the infrared radiation, one may need energy levels as high as 400-600~K in the ground and first excited torsional states to calculate populations reliably at such temperatures.  Too few energy levels will generate truncation errors as the tail of the population distribution is artifically constrained.  Masers can be very sensitive to such errors, as they stem from small differences between nearly equal populations.  When infrared radiation from warm dust is included in the models, it acts to promote methanol molecules to the torsionally excited states, from which they decay back to the ground state in one or more steps.  This disrupts the pattern of population among the ground state levels, giving rise to inversions and hence the class ~II masers.  Although the population which resides in the torsionally excited states is typically very small, the pumping and decay process plays a vital role in determining the populations of the low-lying levels.  It is necessary to include the same number of rotational levels in each accessible torsionally excited state so that all populated ground state levels may participate in the infrared pumping.  Sobolev \& Deguchi (1994) found that both the \vt=1 and \vt=2 levels were essential to the class~II maser pumping.  Here we investigate for the first time the influence of even higher states.

The largest set of energy levels used in our calculations is displayed in Fig.~\ref{fig:levels} for E species methanol.  Details are given in Appendix~A.


\subsection{Rate coefficients for collisions}

In the interstellar gas He atoms constitute about 20 percent of the collision partners, while the remainder is a mixture of ortho- and para-H$_{2}$, with the para ground-state form predominating at low interstellar temperatures.  Lees \& Haque (1974) undertook microwave double-resonance experiments to investigate collisions between E-species methanol with both He and H$_{2}$ at room temperature.  Their results led them to develop the following propensity rules, which have been used in excitation modelling to date in the absence of any other information.  Collision rate coefficients favour the dipole selection rules ($\Delta$J=0,$\pm$1), and decrease as 1/$\mid$$\Delta$J$\mid$ for $\mid$$\Delta$J$\mid>$1, while $\Delta$K=0 transitions are favoured over $\Delta$K=1 transitions by a factor of 4.  Our previous modelling of class~II methanol masers employed these propensity rules in the formulation of Peng \& Whiteoak (1993), extended to the A-species as described in Sobolev \etal\ (1997a), and applied to transitions within \vt=0 and within \vt=1.

Pottage \etal\ (2001) calculated rate coefficients for rotational excitation of methanol by He at temperatures 10 and 20~K, for both A- and E-species methanol up to J=7 in the torsional ground state.  Their results showed trends commensurate with the propensity rules of Lees \& Haque.  In subsequent work (Pottage \etal\ 2002) an improved treatment gave rate coefficients for temperatures between 5 and 200~K, for methanol energy levels up to J=9 in the ground torsional state, plus data at temperature 20~K for the first excited torsional state.  Cross-sections for torsionally inelastic collisions between \vt=0 and \vt=1 were presented by Pottage \etal\ (2004a), again for collision partner He, but in this case the trends did not follow the propensity rules for torsionally elastic transitions.  Rate coefficients for excitation by para-H$_{2}$ were calculated by Pottage \etal\ (2004b), again for temperatures between 5 and 200~K, and for methanol energy levels up to J=9 but in the torsional ground state (\vt=0) only.  As yet there are no calculated rate coefficients for methanol interacting with ortho-H$_{2}$, which remains a potentially significant source of uncertainty.

Excitation modelling requires rate coefficients for collisional transitions between every pair of energy levels included in the calculation.  Since methanol has a huge number of levels which may be populated under interstellar conditions, even the pioneering work of PFD does not provide enough data to fully solve the excitation problem.  It does however provide data for the most important collisional transitions into and out of the energy levels involved in nearly all the observed class~II masers (the exception being the 23.1-GHz $9_{2}-10_{1}\ \rm{A}^{+}$ maser).  In our calculations the \lq missing\rq\ rates have been approximated by extrapolating the given rates in line with the propensity rules, as described in Appendix~B.


\subsection{The Sobolev-Deguchi model}

In the model of Sobolev \& Deguchi (1994) masers develop as the result of radiative pumping by infrared radiation from warm dust, described by dust temperature \Td, filling factor \Wd, and opacity $\taud(\nu / 10^{7})^{2}$, where $\nu$ is the line frequency in MHz.  Radiative transfer is treated in the large velocity gradient (LVG) approximation, augmented by a beaming factor \beam, defined as the ratio of the radial to tangential optical depths, to represent the elongation of the maser region along the line of sight (Castor 1970).  The maser region is treated as a volume of gas of uniform density \nH, kinetic temperature \Tk, and methanol abundance \Xm.  The optical depth is proportional to the specific column density parameter \scdm, defined as the column density of A- or E-species methanol tangential to the line of sight, divided by the line width \DV.  In terms of these parameters the relative abundance of methanol becomes $\Xm=(2 \beam \DV (\scdm)) / (\nH L)$ when the A and E symmetry species are equally abundant, where $L$ is the extent of the maser region along the line of sight.  The model optionally includes free-free emission from an underlying uc\HII\ region, which, if present, will be amplified by the masers.  The \HII\ background continuum has electron temperature \Te\ and free-free turnover frequency \fe, and is geometrically diluted by a factor \WHII.  Here we use \Wd=0.5, \taud=1, \beam=10, \Te=$10^{4}$~K, and \fe=12~GHz throughout, and explore ranges of the parameters \Td, \Tk, \nH, \scdm\ and \WHII\ relevant to class~II methanol masers.


\section{RESULTS}


\subsection{Extra levels}

The number of levels required is a function of the model temperatures, since the gas temperature influences the population distribution among the rotational levels, while the dust temperature mainly governs the distribution between the assorted torsional and vibrational states.  The best way to demonstrate that enough levels have been included in a calculation is to repeat it using more levels and show that the differences are minor.


\begin{figure*}
\centerline{\epsfxsize=18cm\epsfbox{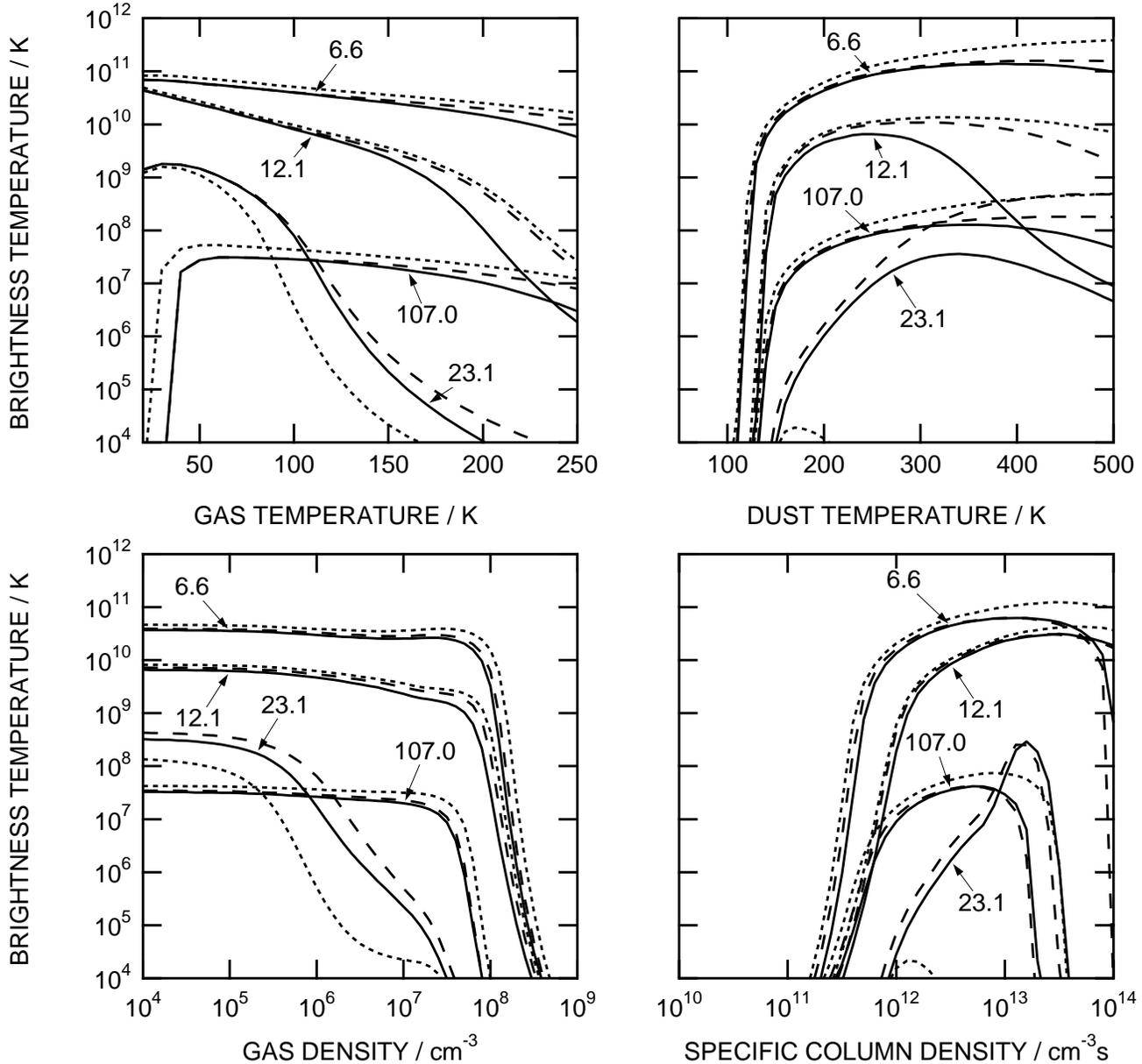}}
\caption{Effects of parameter variation on brightness temperatures for 4 masers, using 3 different data sets.  Solid lines include J=0-22 for \vt=0-4 and \vCO=1, dashed lines include J=0-18 with \vt=0-3, dotted lines include J=0-18 with \vt=0-2.  Fixed parameters are \Tk=150~K, $\Td=175$~K, $\nH=10^{7}$~\cc, $\scdm=10^{12.2}$~\ccs, with no \HII\ region background radiation, and using the new PFD collision data.}
\label{fig:vary_vt}
\end{figure*}


Fig.~\ref{fig:vary_vt} displays model brightness temperatures for 4 selected class~II methanol maser transitions: \six\ at 6.668~GHz, \twelve\ at 12.178~GHz, $9_{2}-10_{1}\ \rm{A}^{+}$ at 23.121~GHz, and $3_{1}-4_{0}\ \rm{A}^{+}$ at 107.013~GHz.  The 6.6- and 12.1-GHz lines are the strongest observed masers, while the 107.0-GHz transition is the most common of the other, weaker masers.  The 23.1-GHz transition involves higher energy levels than any other of the observed class~II masers, and also displays the most sensitivity to the number of levels included among the 23 maser transitions which we have examined in detail.  The calculations displayed make use of the new PFD collision data, but the same trends are seen in calculations using the old propensity rule collision model, so the sensitivity of the 23.1-GHz maser is not due to the extrapolation of the PFD collision rate coefficients above J=9.  The plots show the variation with gas temperature \Tk, dust temperature \Td, gas density \nH, and specific column density of methanol \scdm.  In each case the unvarying parameters are \Tk=150~K, \Td=175~K, $\nH=10^{7}$~\cc, and $\scdm=10^{12.2}$~\ccs.  These models do not include uc\HII\ background continuum radiation.  Fig.~\ref{fig:vary_vt} displays traces for 3 different energy level sets:  the smallest has rotational levels up to J=18 for each of the torsional states \vt=0,1,2 while the largest has rotational levels up to J=22 for \vt=0,1,2,3,4 plus the \vCO=1 vibrationally excited state.  The intermediate set has rotational levels up to J=18 for \vt=0,1,2,3.

It is apparent from Fig.~\ref{fig:vary_vt} that the effects of expanding the energy level set are generally modest.  The smallest set produces masers which follow the same trends as those calculated from the largest set.  The differences are most pronounced at gas temperatures $\Tk>200$~K, at dust temperatures $\Td>300$~K, and at large methanol column densities $\scdm>10^{13}$~\ccs.  A detailed examination including further sets of levels shows that expanding the set of rotational energy levels from J=18 to J=22 generally has a lesser effect than expanding the set of torsional states from \vt=2 to \vt=3.  The intermediate set gives a close approximation to the largest data set except at high dust temperatures.  At $\Td>300$~K the CO-stretch vibration starts to have a significant quenching effect on the masers.  But as we have not included (for lack of available data) the CH$_{3}$ rock vibration which is comparable in energy, we cannot be sure that this regime is reliably modelled even with the largest set.  In the remainder of this paper we use the intermediate data set which includes rotational levels up to J=18 and torsional states up to \vt=3, and we do not investigate dust temperatures above 300~K.  


\subsection{New collisions}


\begin{figure*}
\centerline{\epsfxsize=18cm\epsfbox{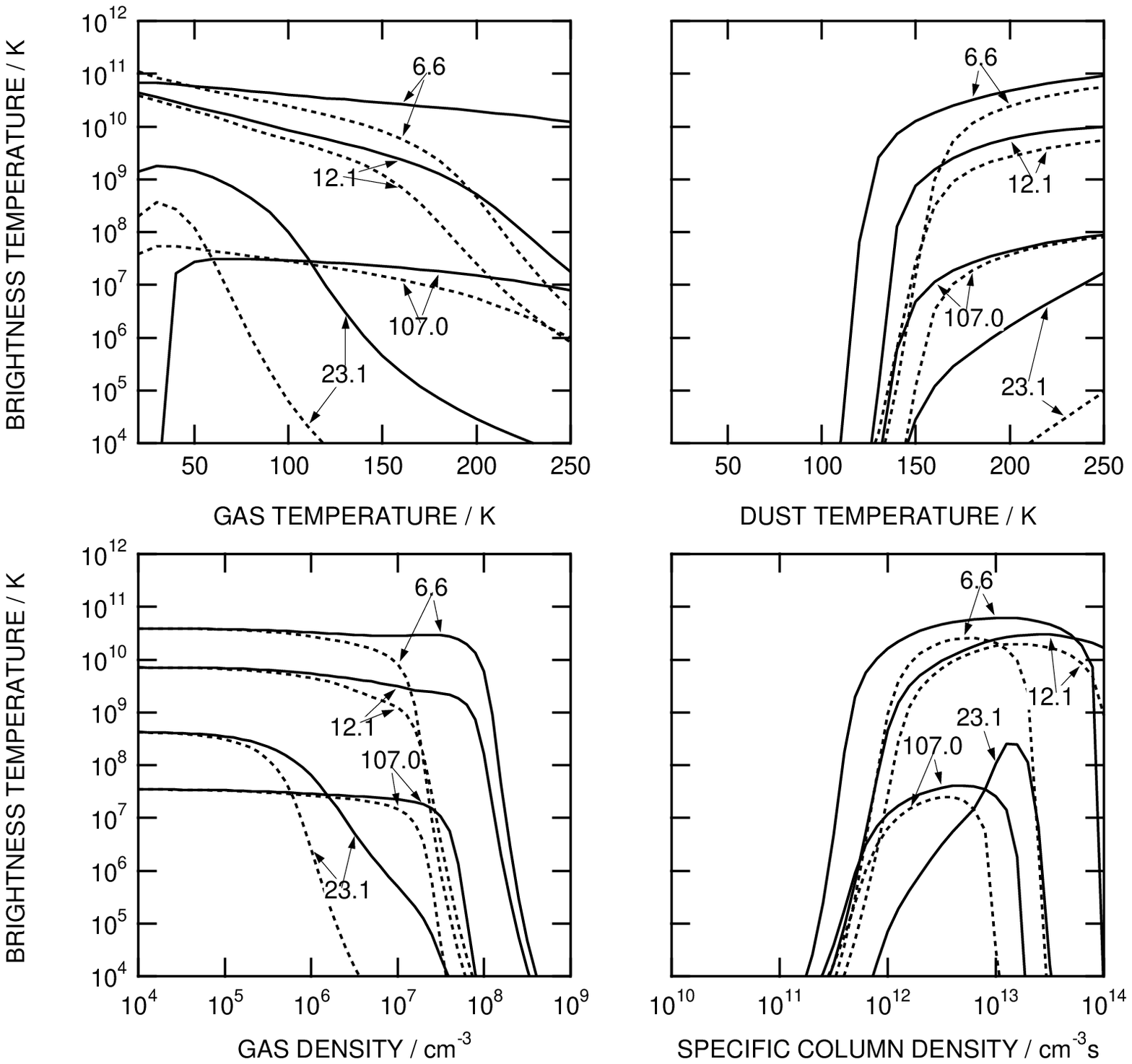}}
\caption{Effects of parameter variation on brightness temperatures for 4 masers, using different collision rate coefficients.  Solid lines use new PFD collision data, dotted lines use propensity rules.  Fixed parameters are \Tk=150~K, $\Td=175$~K, $\nH=10^{7}$~\cc, $\scdm=10^{12.2}$~\ccs, with no \HII\ region background radiation, and using energy levels up to J=18 and \vt=3.}
\label{fig:vary_coll}
\end{figure*}


Fig.~\ref{fig:vary_coll} displays model brightness temperatures for 4 class~II methanol maser transitions, as 4 of the model parameters are varied in the way described in the previous section.  Dotted traces represent the results using the old propensity rule collisions, while the solid traces show the results with the new PFD calculated collision rate coefficients.  Comparing Fig.~\ref{fig:vary_vt} with Fig.~\ref{fig:vary_coll}, it is apparent that the new collision data has much greater impact on the modelling than including more energy levels.  Nevertheless, the behaviour of the masers is recognisably similar with the old and new collision models.  All 4 transitions displayed become masers of similar peak intensity with both collision models, but there are sometimes significant changes in the conditions under which they switch on and off.  The same is true of 19 other observed or potential maser lines which we have examined in detail.

At low gas density ($10^{4}$~\cc) the collision model is irrelevant and the masers are entirely governed by radiative effects.  At high gas density ($10^{9}$~\cc) all the masers are thermally quenched.  The collisions are significant over the intermediate density range, as the masers become progressively quenched.  With the new collisions at \Tk=150~K, all masers extend to higher density than with the old.  The density required for thermalization is increased in some cases by an order of magnitude.  Although some masers are affected more than others, the effects are consistent with an overall reduction in the collision cross-section with the new model.  At \Tk=150~K the masers appear at a lower dust temperature threshold, and extend over a wider range of specific column density, with the new collision model.  These effects are seen when the gas density is moderately high (here $\nH=10^{7}$~\cc), meaning that collisions contribute significantly to the balance of effects in the maser pumping.  Both these effects are consistent with an overall reduction in collision cross-section, so that the radiative pumping becomes more influential.

The picture changes as the gas temperature varies.  At gas temperatures above 100~K the new collisions produce enhanced maser intensity at high density, as described above.  At gas temperatures below 100~K it is not so clear-cut.  For example, at \Tk=30~K some masers extend to higher density with the new collisions, while others are thermalized at lower density than before.  Thus the significant collision processes (whatever these may be) have become more probable for some transitions and less probable for others.  Corresponding effects appear in the behaviour as specific column density and dust temperature are varied for \Tk=30~K.  The net effect is that at low gas temperatures some of the weaker maser lines are enhanced while others are diminished.


\subsection{Limits on most common maser sources}


\begin{figure}
\centerline{\epsfxsize=8.5cm\epsfbox{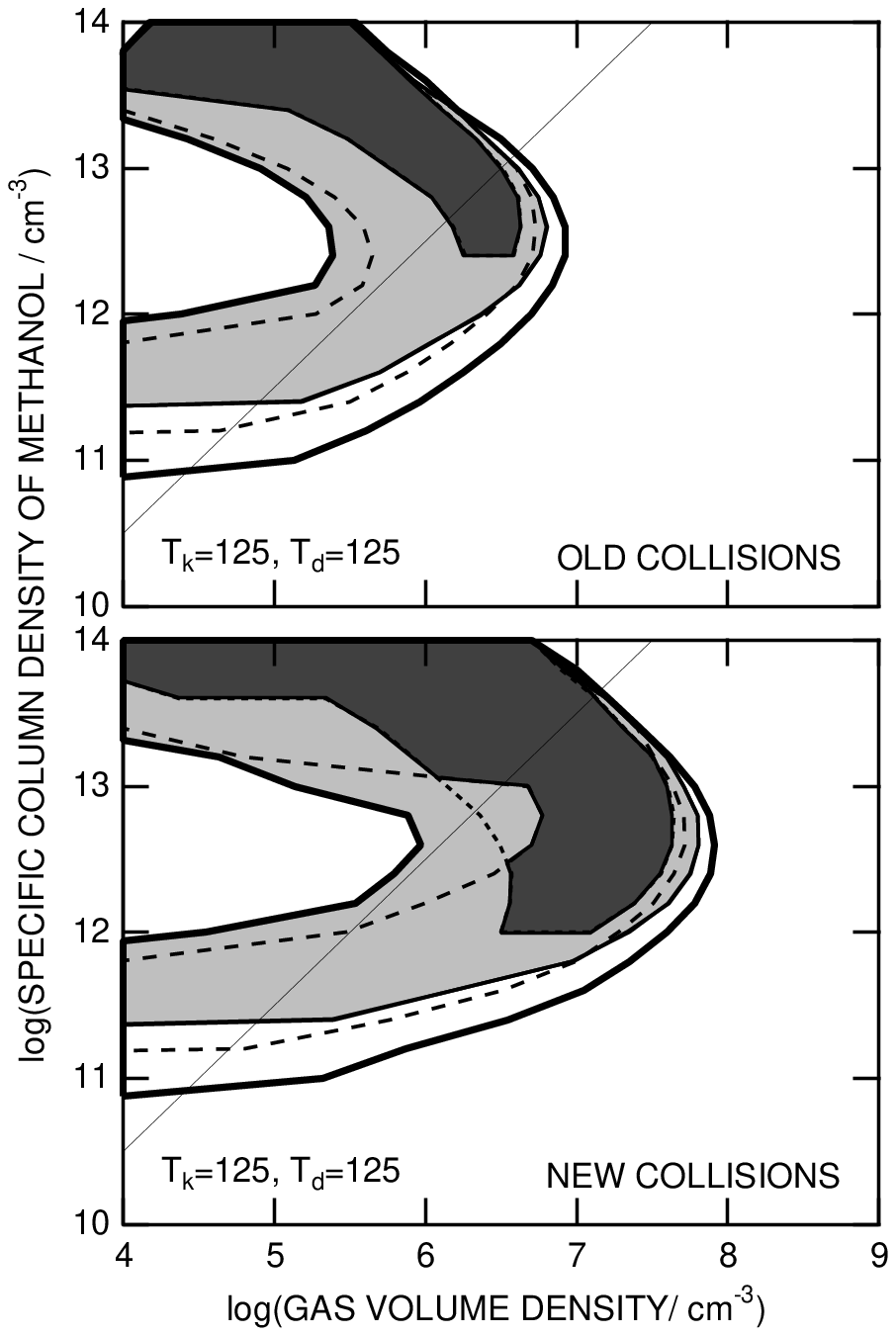}}
\caption{Heavy and dashed contours delimit the model parameter combinations consistent with the strongest and weakest upper limits on maser emission from survey at 23.1~GHz, while light and dotted contours do the same for 107.0~GHz.  The shaded ares represent conditions satisfying upper limits for both transitions, for sources with masers detected at 6.6~GHz, but not at 23.1 and 107.0~GHz.  The line $\log(\scdm) < \log(\nH) + 6.5$ distinguishes plausible model conditions (lower right) from those requiring unrealistically large maser path lengths or methanol abundance (upper left).  Top panel is from Cragg \etal\ (2004) using the propensity rules for collisions, while bottom panel is based on the PFD collision rate coefficients.  }
\label{fig:limits}
\end{figure}


The majority of class~II methanol maser sources have been observed only at 6.6~GHz.  Surveys at 107.0 and 23.1~GHz have demonstrated that the weaker maser transitions are rarely detected at 6.6-GHz maser sites (Val'tts \etal\ 1995, 1999; Caswell \etal\ 2000; Minier \& Booth 2002; Cragg \etal\ 2004).  It is evident from Fig.~\ref{fig:vary_coll} that the model predicts emission from the weaker masers simultaneously with the 6.6-GHz maser under many combinations of parameters.  However, the 6.6-GHz maser is the brightest and the most readily excited under the conditions investigated.  Sources in which only the 6.6-GHz maser is detected may experience conditions near the edges of the maser regime, or may have maser emission in the weaker transitions which is below the detectable limit.  In Cragg \etal\ (2004) we used the upper limits on maser emission at 107.0 and 23.1~GHz to investigate conditions in 28 class~II maser sources detected at 6.6~GHz, by using these constraints to define models which give ratios compatible with the observations.  We reported model calculations for gas temperatures \Tk=25, 75, 125, 175~K, dust temperatures \Td=125, 175, 225, 275~K, gas densities in the range $10^{4} \le \nH \le 10^{9}$~\cc, and methanol specific column densities in the range $10^{10} \le \scdm \le 10^{14}$~\ccs.  The background uc\HII\ radiation was omitted in this case, since less than half the class~II methanol maser sources have associated ultra-compact \HII\ regions, and it is suggested that the majority of masers accompany the very early stages of high-mass star formation, before the development of a detectable uc\HII\ continuum (Ellingsen \etal\ 1996).  These calculations used the propensity rules to define collisional excitation rates, and  we have now repeated them using the new PFD calculated collision rate coefficients.  While the new results may be considered more reliable, note that the extrapolation of the PFD data for J$>$9 leaves some uncertainty over the behaviour of the 23.1-GHz $9_{2}-10_{1}\ \rm{A}^{+}$ maser.

We find that as a general rule the 6.6-GHz maser extends to greater values of gas density with the new collisions than with the old, sometimes as high as $10^{9}$~\cc, consistent with the examples in Fig.~\ref{fig:vary_coll}.  The difference is greatest when \Tk=175~K and \Td=125~K, where the 6.6-GHz masers extend to more than an order of magnitude greater density than previously.  This suggests that relevant collisional processes which act to quench the masers have become less probable with the new rate coefficents.  The same trend can be seen for the 23.1-GHz masers, whereas for the 107.0-GHz masers the effect is less pronounced at high gas temperatures and sometimes reversed at low gas temperatures.  Consistent with this pattern, the masers often extend to slightly higher values of \scdm\ and switch on at lower thresholds of \Td\ than in previous models.  The analysis of Cragg \etal\ (2004) was based on ratios of brightness temperature in the various transitions, since absolute values cannot be determined from observed flux densities without knowledge of the source size.  Repeating this analysis to define parameter combinations consistent with observed limits on maser emission, we obtain similar results to before, but generally shifted to slightly higher values of gas density, as the following example illustrates.

Fig.~\ref{fig:limits} compares the range of \nH\ and \scdm\ compatible with the observed upper limits at \Tk=125~K and \Td=125~K with the old and new collision models.  Dark and light shaded regions show parameter ranges in which brightness temperatures of the 23.1- and 107.0-GHz masers relative to the 6.6-GHz maser are below the strongest and weakest limits (respectively) set by the survey data.  The results with the new collisions can be summarized as a shift to higher gas density by about a factor of 10, so that somewhat higher gas densities are now allowed by the survey limits, together with some relaxation of the limit on methanol column density.  A shift in the same direction is seen with other \Tk\ and \Td\ combinations.  The change is particularly apparent when \Tk$\geq$\Td\ (as in Fig.~\ref{fig:limits}), because the restriction $\log(\scdm) < \log(\nH) + 6.5$ (which ensures reasonable values of the methanol abundance \Xm\ and maser path length $L$) now leaves a more substantial range of plausible models.


\subsection{W3(OH)}


\begin{table*}

\caption{Comparison between maser flux densities observed towards W3(OH) and 4 model calculations, following Sutton \etal\ (2001).  Observational references (a) Sutton \etal\ (2001), (b) Menten \etal\ (1988) , (c) Wilson \etal\ (1993) , (d) Haschick, Baan \& Menten (1989) , (e) Slysh, Kalenskii \& Val'tts (1995) , (f) Menten \etal\ (1992) , (g) Moscadelli \etal\ (1999) , (h) Wilson \etal\ (1985) .} 

\label{tab:w3oh}

\centerline{
\begin{tabular}{rrrrrrrrrrrr}\hline\\                          
&&&   \multicolumn{2}{c}{Model B}   & \multicolumn{2}{c}{Model C}   & \multicolumn{2}{c}{Model D}   & \multicolumn{2}{c}{Model E} & \\
\multicolumn{3}{r}{collisions}  & \multicolumn{2}{c}{OLD}   & \multicolumn{2}{c}{NEW}   & \multicolumn{2}{c}{NEW}   & \multicolumn{2}{c}{NEW} & \\
\multicolumn{3}{r}{\Tk}  &   \multicolumn{2}{c}{150}   & \multicolumn{2}{c}{150}   & \multicolumn{2}{c}{150}   & \multicolumn{2}{c}{145} & \\
\multicolumn{3}{r}{\Td}  & \multicolumn{2}{c}{175}   & \multicolumn{2}{c}{175}   & \multicolumn{2}{c}{175}   & \multicolumn{2}{c}{145} & \\
\multicolumn{3}{r}{log(\nH)}  & \multicolumn{2}{c}{7.0}   & \multicolumn{2}{c}{7.0}   & \multicolumn{2}{c}{7.5}   & \multicolumn{2}{c}{7.0} & \\
\multicolumn{3}{r}{log(\scdm)}   & \multicolumn{2}{c}{12.0}   & \multicolumn{2}{c}{12.0}   & \multicolumn{2}{c}{12.0}   & \multicolumn{2}{c}{12.0} & \\
\multicolumn{3}{r}{\WHII}   & \multicolumn{2}{c}{0.2}   & \multicolumn{2}{c}{0.2}   & \multicolumn{2}{c}{0.2}   & \multicolumn{2}{c}{0.2} & \\
\hline\\                          
\multicolumn{1}{c}{Transition}   & \multicolumn{1}{c}{Frequency}  &   \multicolumn{1}{c}{$S_{obs}$}  & \multicolumn{1}{c}{$T_{calc}$}  & \multicolumn{1}{c}{$S_{calc}$}  & \multicolumn{1}{c}{$T_{calc}$}  & \multicolumn{1}{c}{$S_{calc}$}  & \multicolumn{1}{c}{$T_{calc}$}  & \multicolumn{1}{c}{$S_{calc}$}  & \multicolumn{1}{c}{$T_{calc}$}  & \multicolumn{1}{c}{$S_{calc}$} & ref\\
 &   \multicolumn{1}{c}{(MHz)} &   \multicolumn{1}{c}{(Jy)} & \multicolumn{1}{c}{(K)} & \multicolumn{1}{c}{(Jy)} & \multicolumn{1}{c}{(K)} & \multicolumn{1}{c}{(Jy)} & \multicolumn{1}{c}{(K)} & \multicolumn{1}{c}{(Jy)} & \multicolumn{1}{c}{(K)} & \multicolumn{1}{c}{(Jy)} & \\
\hline\\                          
\multicolumn{12}{l}{BIMA observations, included in the fit} \\
7(2) -- 8(1) A-- & 80993.3 & $<$ 0.3 &  9.53E+02 & 0.0 &  5.56E+04 & 1.8 & 2.36E+03 & 0.2 & 7.92E+02 & 0.2 & (a) \\
13(--3) -- 14(--2) E & 84423.7 & $<$ 0.3 &  1.25E+02 & 0.0 &  2.89E+04 & 1.0 & 7.13E+03 & 0.5 & 3.01E+02 & 0.1 & (a) \\
6(--2) -- 7(--1) E & 85568.1 & $<$ 0.7 &  --1.40E+02 & 0.0 &  --1.02E+02 & 0.0 & 2.38E+00 & 0.0 & --1.33E+02 & 0.0 & (a) \\
7(2) -- 6(3) A-- & 86615.6 &  6.7 &  1.99E+05 & 8.0 &  1.02E+06 & 37.0 & 1.28E+05 & 9.7 & 3.93E+04 & 10.2 & (a) \\
7(2) -- 6(3) A+ & 86902.9 &  7.2 &  1.82E+05 & 7.3 &  9.48E+05 & 34.6 & 1.15E+05 & 8.8 & 3.08E+04 & 8.0 & (a) \\
8(3) -- 9(2) E & 94541.8 & $<$ 0.4 &  --5.40E+01 & 0.0 &  2.59E+02 & 0.0 & 2.57E+02 & 0.0 & --5.49E+01 & 0.0 & (a) \\
3(1) -- 4(0) A+ & 107013.8 &  72.0 &  1.18E+06 & 72.0 &  1.30E+06 & 72.0 & 6.23E+05 & 72.0 & 1.82E+05 & 72.0 & (a) \\
0(0) -- 1(--1) E & 108893.9 & $<$ 0.6 &  8.09E+03 & 0.5 &  5.66E+04 & 3.2 & 4.25E+04 & 5.1 & 3.00E+03 & 1.2 & (a) \\
7(2) -- 8(1) A+ & 111289.6 & $<$ 1.0 &  2.86E+02 & 0.0 &  5.19E+03 & 0.3 & 9.50E+02 & 0.1 & 3.40E+02 & 0.1 & (a) \\
\multicolumn{12}{l}{previous observations,  not included in the fit} \\
9(2) -- 10(1) A+ & 23121.0 &  9.5 &  3.42E+03 & 0.0 &  1.35E+06 & 3.5 & 1.44E+05 & 0.8 & 3.12E+04 & 0.6 & (b) \\
4(0) -- 3(1) E & 28316.0 &  --0.6 &  --1.61E+03 & 0.0 &  --1.63E+03 & 0.0 & --1.62E+03 & 0.0 & --1.61E+03 & 0.0 & (c) \\
8(2) -- 9(1) A-- & 28970.0 &  13.8 &  1.38E+04 & 0.1 &  2.22E+06 & 9.0 & 1.75E+05 & 1.5 & 7.39E+04 & 2.1 & (c) \\
6(2) -- 5(3) A-- & 38293.3 &  14.0 &  2.00E+05 & 1.6 &  3.19E+06 & 22.6 & 3.26E+05 & 4.8 & 1.38E+05 & 7.0 & (d) \\
6(2) -- 5(3) A+ & 38452.7 &  9.0 &  1.78E+05 & 1.4 &  2.89E+06 & 20.7 & 3.02E+05 & 4.5 & 1.14E+05 & 5.8 & (d) \\
8(0) -- 8(--1) E & 156488.9 &  27.9 &  2.40E+06 & 313.1 &  2.18E+06 & 258.2 & 1.39E+06 & 343.5 & 8.12E+04 & 68.7 & (e) \\
2(1) -- 3(0) A+ & 156602.3 & $<$ 9.0 &  2.73E+05 & 35.7 &  2.56E+05 & 30.4 & 1.77E+04 & 4.4 & 2.15E+03 & 1.8 & (e) \\
7(0) -- 7(--1) E & 156828.5 &  32.7 &  2.39E+06 & 313.2 &  2.21E+06 & 262.9 & 1.37E+06 & 340.0 & 6.42E+04 & 54.5 & (e) \\
5(0) -- 5(--1) E & 157179.0 &  24.2 &  2.43E+06 & 319.9 &  1.92E+06 & 229.4 & 1.16E+06 & 289.2 & 5.01E+04 & 42.8 & (e) \\
4(0) -- 4(--1) E & 157246.0 &  13.5 &  1.42E+06 & 187.1 &  1.52E+06 & 181.8 & 8.07E+05 & 201.4 & 1.79E+04 & 15.3 & (e) \\
\multicolumn{12}{l}{other masers with complex profiles,  not included in the fit} \\
5(1) -- 6(0) A+ & 6668.5 &  756.0 &  2.31E+08 & 54.7 &  3.44E+08 & 74.0 & 3.81E+08 & 171.0 & 1.38E+08 & 212.0 & (f) \\
2(0) -- 3(--1) E & 12178.6 &  10.0 &  4.08E+07 & 32.2 &  5.22E+07 & 37.4 & 4.91E+07 & 73.5 & 1.18E+07 & 60.5 & (g) \\
2(1) -- 3(0) E & 19967.4 &  55.0 &  3.33E+04 & 0.1 &  4.98E+05 & 1.0 & 5.14E+04 & 0.2 & 5.38E+04 & 0.7 & (h) \\
7(--2) -- 8(--1) E & 37703.7 &  4.3 &  --9.26E+02 & 0.0 &  --8.93E+02 & 0.0 & --3.75E+02 & 0.0 & --9.29E+02 & 0.0 & (d) \\
\hline\\                          
\end{tabular}}                          

\end{table*}


There are a few exceptional class~II methanol maser sources in which masers have been detected at many frequencies.  Although they have complex spectra at 6.6 and 12.1~GHz, with many maser components spaced over a narrow range of velocities, usually only one or two components are evident in the weaker maser lines.  When the velocities match it suggests that the masers at different frequencies are simultaneously excited in the same volume of gas.  Interferometric observations have confirmed that the 6.6- and 12.1-GHz emission in some sources is spatially coincident to within a few milliarcseconds (Menten \etal\ 1992; Norris \etal\ 1993; Minier, Booth \& Conway 2000).  In the models such simultaneous excitation is readily achieved (in fact it is difficult to find conditions which would explain emission in one transition alone), but the combinations of masers which appear together are sensitive to the conditions.  Thus, when masers are detected in several transitions, they provide a sensitive probe of the local conditions, so long as they coincide.  Upper limits on maser nondetections can also provide constraints on the physical conditions, and with fewer assumptions since the limits apply to all maser sites within the beam.

In Sutton \etal\ (2001) we exploited this method to estimate physical parameters for the maser region in W3(OH), based on BIMA observations of maser spikes at 86.6, 86.9 and 107.0~GHz at velocity -43.1 \kms, which were found to be located near the northern edge of the uc\HII\ region.  No corresponding spike emission was obtained for the 80.9, 84.4, 85.5, 94.5, 108.8, or 111.2-GHz maser candidates.  While maser emission at several other methanol frequencies has been detected at the same velocity (Table~\ref{tab:w3oh}), there are at least 2 distinct maser regions in W3(OH) emitting at this velocity (Menten \etal\ 1992), and so it is not readily apparent from single-dish observations whether the remaining masers coincide.  Indeed, the 23.1-GHz maser peak emission is from the southern edge of the uc\HII\ (Menten \etal\ 1988), unlike the 107.0-GHz maser (Sutton \etal\ 2001), suggesting that local conditions at these two sites are rather different.

We have undertaken further model calculations based on the new PFD collision rate coefficients to compare with the W3(OH) observations.  These models include continuum radiation from an underlying uc\HII\ region with \WHII=0.2.  The results are summarized in Table~\ref{tab:w3oh}, which includes data (maser detections or upper limits) for the 9 transitions observed by Sutton \etal\ and included in the fit, 10 other transitions with comparable maser profiles (but not necessarily emanating from the same site), and 4 further masers with complex multi-component profiles.  The observed flux densities are compared with the predictions of 4 models.  For each model, calculated brightness temperatures are converted to flux densities based on a common source size which matches the observed flux density at 107.0~GHz.

Model~B reproduces the calculations of Sutton \etal\ (2001) using the propensity rule collisions, with parameters \Tk=150, \Td=175, log(\nH)=7.0, log(\scdm)=12.0.  This gives a good fit to the 86-GHz masers, and meets the upper limits at 80.9, 84.4, 85.5, 94.5, 108.8 and 111.2~GHz.  Model~C has the same parameters but uses the new collisions.  The 86-GHz masers are far too strong, and the limits at 80.9, 84.4 and 108.8~GHz are not satisfied.  Model~D also uses the new collisions, but has a higher gas density, log(\nH)=7.5.  It gives a good fit to the 86-GHz intensities, but still fails the upper limits at 84.4 and 108.8~GHz.  This is consistent with the conclusions of the last section, where (to a first approximation, and at high gas temperatures) the effect of the new collisions is comparable to a density scaling, such that higher densities are now required.  Model~E again uses the new collisions, with lower temperatures \Tk=\Td=145, and log(\nH)=7.0, log(\scdm)=12.0.  The equal dust and gas temperatures generate masers of lower brightness temperature than the other models shown, and so a correspondingly larger source size is required to match the observed flux densities.  It gives a good fit to the 86-GHz intensities, and meets all the upper limits except 108.8~GHz, where it is an improvement on Models C and D.  Note that this model makes predictions more consistent than the others with the 157-GHz J$_{0}$-J$_{-1}$~E maser observations.  These observations are not included in the fit, because it is not known whether they emanate from the same northern source, but in either case the other models overestimate them by at least an order of magnitude.  Thus a comparable, or perhaps better, account of the observations can be obtained with the new collisions, under conditions which are not vastly different.


\subsection{Maser Candidates}


\begin{table*}

\caption{Class II methanol maser candidate lines, which have brightness temperature $>10^{4}$~K in one or more of the 8 models presented.  All model calculations employ the new collision data of PFD.  Blank entries signify brightness temperatures $<10^{3}$~K, while brightness temperatures $>10^{6}$~K are in bold.  Entries which meet the specified thresholds, but represent quasi-thermal rather than maser emission, are marked with an asterisk.} 

\label{tab:predictions}
\centerline{
\begin{tabular}{cccrrrrrrrrr}\hline\\                          
&&&                    MODEL      &           1 &    2 &    3 &    4 &    5 &    6 &    7 &    8\\
&&&                    \Tk        &         150 &  150 &   30 &  125 &  125 &  145 &   50 &  120\\
&&&                    \Td        &         175 &  175 &  175 &  175 &  175 &  145 &  125 &  110\\
&&&                    log(\nH)   &         4.0 &  7.0 &  7.0 &  8.0 &  8.0 &  7.0 &  5.9 &  6.0\\
&&&                    log(\scdm) &        11.5 & 11.5 & 11.5 & 12.0 & 12.0 & 12.0 & 12.0 & 11.5\\
&&&                    \WHII      &       0.002 &0.002 &0.002 &   -- &0.002 &  0.2 &0.002 &0.002\\
\\
\vt& Upper Level &  Lower Level  &  \multicolumn{1}{c}{Frequency/MHz}   &     \multicolumn{8}{c}{log(BRIGHTNESS TEMPERATURE /K)}\\
\\
 0 &   6(2)  A+  &   6(2)  A--  &     158.8  &      &     &   3.4 &      &      &      &  4.3 &      \\ 
 0 &   8(2)  A+  &   8(2)  A--  &     476.1  &      &     &   3.2 &      &      &      &  4.1 &      \\ 
 0 &  11(2)  E   &  10(3)  E    &    2926.0  &      &  3.3&       &      &      &  4.9 &      & \bf{6.3} \\ 
 0 &   3(1)  A-- &   3(1)  A+   &    5005.3  &      &     &   4.1 &      &      &      &  3.6 &      \\ 
 0 &   5(1)  A+  &   6(0)  A+   &    6668.5  & \bf{9.9} & \bf{9.3}& \bf{10.5} & \bf{9.2} &\bf{10.3} &\bf{10.0} &\bf{10.0} & \bf{6.3} \\ 
 0 &  11(-3)  E  &  10(-4)  E   &    7283.4  &  4.6 &  4.9&       &      &  5.9 & \bf{6.3} &      &      \\ 
 0 &  12(4)  A-- &  13(3)  A--  &    7682.3  &  4.3 &  3.3&   3.8 &      &  4.1 &  3.9 &  4.6 &      \\ 
 0 &  12(4)  A+  &  13(3)  A+   &    7830.9  &  4.3 &  3.4&   3.8 &      &  4.1 &  3.9 &  4.5 &      \\ 
 0 &   4(1)  A-- &   4(1)  A+   &    8341.6  &      &     &   4.6 &      &      &  3.2 &  4.6 &      \\ 
 0 &   9(-1)  E  &   8(-2)  E   &    9936.2  &      &  3.9&       &      &      & \bf{6.4} &  5.5 & \bf{8.2} \\ 
 0 &   2(0)  E   &   3(-1)  E   &   12178.6  & \bf{8.5} & \bf{7.1}& \bf{10.0} & \bf{7.1} & \bf{9.2} & \bf{8.6} & \bf{9.5} &      \\ 
 0 &   5(1)  A-- &   5(1)  A+   &   12511.2  &  3.3 &     &   4.2 &      &      &  3.4 &  4.5 &      \\ 
 0 &   2(1)  E   &   3(0)  E    &   19967.4  &  5.4 &  4.0&  \bf{8.7} &      &  3.6 &  4.7 & \bf{8.7} &  3.6 \\ 
 0 &  11(1)  A+  &  10(2)  A+   &   20171.1  &      &     &       &      &      &      &      &  4.3 \\ 
 0 &  16(-4)  E  &  15(-5)  E   &   20908.8  &  3.8 &  4.1&       &      &  3.4 &  5.0 &  3.6 &  3.3 \\ 
 1 &  10(1)  A+  &  11(2)  A+   &   20970.7  &  3.7 &  3.4&   3.7 &      &  4.0 &  3.7 &  4.1 &      \\ 
 0 &   9(2)  A+  &  10(1)  A+   &   23121.0  & \bf{6.0} &  4.0&  \bf{7.7} &      &  3.9 &  4.5 & \bf{8.6} &      \\ 
 0 &  10(1)  A-- &   9(2)  A--  &   23444.8  &      &     &       &      &      &      &      &  4.7 \\ 
 0 &   4(2)  E   &   4(1)  E    &   24933.5  &      &     &       &      &      &      &      &  4.1 \\ 
 0 &   5(2)  E   &   5(1)  E    &   24959.1  &      &     &       &      &      &      &      &  4.5 \\ 
 0 &   6(2)  E   &   6(1)  E    &   25018.1  &      &     &       &      &      &      &      &  4.7 \\ 
 0 &   7(2)  E   &   7(1)  E    &   25124.9  &      &     &       &      &      &      &      &  4.9 \\ 
 0 &   8(2)  E   &   8(1)  E    &   25294.4  &      &     &       &      &      &      &      &  4.9 \\ 
 0 &   9(2)  E   &   9(1)  E    &   25541.4  &      &     &       &      &      &      &      &  4.9 \\ 
 0 &  10(2)  E   &  10(1)  E    &   25878.2  &      &     &       &      &      &      &      &  4.9 \\ 
 0 &  11(2)  E   &  11(1)  E    &   26313.1  &      &     &       &      &      &      &      &  4.8 \\ 
 0 &  12(2)  E   &  12(1)  E    &   26847.2  &      &     &       &      &      &      &      &  4.6 \\ 
 0 &  13(2)  E   &  13(1)  E    &   27472.5  &      &     &       &      &      &      &      &  4.4 \\ 
 0 &   8(2)  A-- &   9(1)  A--  &   28970.0  & \bf{6.4} &  4.0&  \bf{8.3} &      &  3.7 &  4.9 & \bf{8.6} &      \\ 
 0 &  14(-3)  E  &  15(-2)  E   &   34236.8  &  3.7 &     &   3.2 &      &  3.9 &      &  4.0 &      \\ 
 0 &   7(-2)  E  &   8(-1)  E   &   37703.7  &      &     &  \bf{7.0} &      &  3.4 &      &      &      \\ 
 0 &   6(2)  A-- &   5(3)  A--  &   38293.3  &  4.7 &  3.9&   5.4 &      &  4.2 &  5.2 & \bf{6.8} &      \\ 
 0 &   6(2)  A+  &   5(3)  A+   &   38452.7  &  4.7 &  3.8&   5.5 &      &  4.2 &  5.1 & \bf{7.2} &      \\ 
 1 &   2(0)  E   &   3(1)  E    &   44955.8  &      &     &   4.2 &      &      &      &  3.7 &      \\ 
 0 &   9(3)  E   &  10(2)  E    &   45843.6  &  3.2 &     &   4.9 &      &      &      &  4.0 &      \\ 
 0 &  12(2)  E   &  11(3)  E    &   51759.5  &      &     &       &      &      &  3.1 &      &  4.5 \\ 
 0 &  12(-3)  E  &  11(-4)  E   &   55673.9  &      &  3.3&       &  3.1 &  4.0 &  4.2 &      &      \\ 
 0 &   4(1)  A+  &   5(0)  A+   &   57032.9  & \bf{7.8} & \bf{6.7}&  \bf{8.4} &  5.8 & \bf{6.2} & \bf{7.7} & \bf{7.7} &  3.1 \\ 
 0 &  10(-1)  E  &   9(-2)  E   &   57292.9  &      &     &       &      &      &  4.0 &      & \bf{6.2} \\ 
 0 &   1(0)  E   &   2(-1)  E   &   60531.5  &  5.9 &  4.9&  \bf{7.5} &  5.3 & \bf{6.0} &  5.1 &  4.8 &      \\ 
 0 &   8(2)  A+  &   9(1)  A+   &   66947.9  &  5.0 &  3.1&  \bf{7.3} &      &      &  3.3 & \bf{7.6} &      \\ 
 0 &   1(1)  E   &   2(0)  E    &   68305.7  &  3.6 &     &  \bf{7.2} &      &      &  3.6 & \bf{7.7} &      \\ 
 0 &   7(2)  A-- &   8(1)  A--  &   80993.3  &  5.1 &     &  \bf{7.6} &      &      &      & \bf{7.5} &      \\ 
 0 &   6(-2)  E  &   7(-1)  E   &   85568.1  &      &     &   5.6 &      &      &      &      &      \\ 
 0 &   7(2)  A-- &   6(3)  A--  &   86615.6  &  4.2 &  3.3&   4.0 &  3.1 &  3.3 &  4.6 &  4.3 &      \\ 
 0 &   7(2)  A+  &   6(3)  A+   &   86902.9  &  4.0 &  3.3&   3.9 &  3.1 &  3.3 &  4.5 &  4.5 &      \\ 
 0 &   8(3)  E   &   9(2)  E    &   94541.8  &      &     &   4.5 &      &      &      &      &      \\ 
 0 &  11(-1)  E  &  10(-2)  E   &  104300.4  &      &     &       &      &      &  3.1 &      &  4.9 \\ 
 0 &   3(1)  A+  &   4(0)  A+   &  107013.8  & \bf{6.6} &  5.0&  \bf{7.1} &* 3.1 &* 3.0 &  5.8 & \bf{6.1} &      \\ 
 0 &   0(0)  E   &   1(-1)  E   &  108893.9  &  4.1 &  3.7&       &  4.0 &  4.2 &  3.6 &      &      \\ 
 0 &   7(2)  A+  &   8(1)  A+   &  111289.6  &  4.3 &     &  \bf{7.2} &      &      &      & \bf{6.9} &      \\ 
 0 &   3(1)  E   &   2(2)  E    &  120197.5  &  3.1 &     &   4.0 &      &      &  3.1 &  4.1 &      \\ 
\end{tabular}}                          

\end{table*}

\begin{table*}

\contcaption{}

\centerline{
\begin{tabular}{cccrrrrrrrrr}\hline\\                          
&&&                    MODEL      &           1 &    2 &    3 &    4 &    5 &    6 &    7 &    8\\
\\
\vt& Upper Level &  Lower Level  &  \multicolumn{1}{c}{Frequency/MHz}   &     \multicolumn{8}{c}{log(BRIGHTNESS TEMPERATURE /K)}\\
\\
 0 &   6(2)  A-- &   7(1)  A--  &  132621.9  &  4.2 &     &  \bf{6.9} &      &      &      & \bf{6.7} &      \\ 
 0 &   8(2)  A-- &   7(3)  A--  &  134896.9  &  3.6 &  3.1&       &      &      &  4.2 &      &      \\ 
 0 &   8(2)  A+  &   7(3)  A+   &  135376.9  &  3.6 &  3.1&       &      &      &  4.2 &      &      \\ 
 0 &   7(3)  E   &   8(2)  E    &  143169.5  &      &     &   4.2 &      &      &      &      &      \\ 
 0 &  11(0)  E   &  11(-1)  E   &  154425.8  &  4.7 &  3.8&   3.0 &* 3.1 &* 3.1 &  3.9 &      &      \\ 
 0 &  10(0)  E   &  10(-1)  E   &  155320.8  &  5.5 &  4.3&   3.6 &* 3.2 &* 3.1 &  4.4 &      &      \\ 
 0 &   9(0)  E   &   9(-1)  E   &  155997.5  & \bf{6.0} &  4.8&   4.6 &* 3.2 &* 3.2 &  4.7 &* 3.3 &      \\ 
 0 &   6(2)  A+  &   7(1)  A+   &  156127.7  &  3.5 &     &  \bf{6.6} &      &      &      & \bf{6.1} &      \\ 
 0 &   8(0)  E   &   8(-1)  E   &  156488.9  & \bf{6.2} &  5.3&   5.5 &* 3.5 &* 3.4 &  5.1 &  4.2 &      \\ 
 0 &   2(1)  A+  &   3(0)  A+   &  156602.3  &  5.0 &  3.7&   4.8 &      &      &  3.3 &      &      \\ 
 0 &   7(0)  E   &   7(-1)  E   &  156828.5  & \bf{6.3} &  5.5&  \bf{6.0} &* 3.3 &* 3.3 &  5.0 &  4.7 &      \\ 
 0 &   6(0)  E   &   6(-1)  E   &  157048.6  & \bf{6.4} &  5.6&  \bf{6.3} &* 3.4 &* 3.3 &  5.0 &  4.9 &      \\ 
 0 &   5(0)  E   &   5(-1)  E   &  157179.0  & \bf{6.4} &  5.7&  \bf{6.3} &* 3.3 &* 3.3 &  4.9 &  4.7 &      \\ 
 0 &   4(0)  E   &   4(-1)  E   &  157246.0  & \bf{6.3} &  5.5&  \bf{6.3} &* 3.3 &* 3.2 &  4.4 &  4.2 &      \\ 
 0 &   1(0)  E   &   1(-1)  E   &  157270.8  &  4.1 &  3.7&       &      &      &      &      &      \\ 
 0 &   3(0)  E   &   3(-1)  E   &  157272.3  & \bf{6.0} &  5.2&       &* 3.1 &      &  3.6 &      &      \\ 
 0 &   2(0)  E   &   2(-1)  E   &  157276.0  &  5.3 &  4.5&       &      &      &      &      &      \\ 
 0 &   1(1)  E   &   1(0)  E    &  165050.2  &  4.2 &     &  \bf{6.3} &      &      &  3.3 &  5.1 &      \\ 
 0 &   2(1)  E   &   2(0)  E    &  165061.2  &  4.2 &     &  \bf{6.3} &      &      &      &* 3.3 &      \\ 
 0 &   3(1)  E   &   3(0)  E    &  165099.3  &  4.3 &     &       &      &      &      &      &      \\ 
 0 &   4(1)  E   &   4(0)  E    &  165190.5  &  4.7 &     &       &      &      &      &      &      \\ 
 0 &   5(1)  E   &   5(0)  E    &  165369.4  &  4.7 &     &       &      &      &      &      &      \\ 
 0 &   6(1)  E   &   6(0)  E    &  165678.7  &  4.6 &     &       &      &      &      &      &      \\ 
 0 &   7(1)  E   &   7(0)  E    &  166169.2  &  4.2 &     &       &      &      &      &      &      \\ 
 0 &   4(1)  E   &   3(2)  E    &  168577.9  &  3.8 &  3.0&   4.3 &      &      &  3.5 &  4.8 &      \\ 
 0 &  11(-3)  E  &  12(-2)  E   &  183720.1  &  3.1 &     &   3.3 &      &      &      &  4.2 &      \\ 
 0 &   5(2)  A-- &   6(1)  A--  &  183852.8  &  3.2 &     &  \bf{6.2} &      &      &      &  5.9 &      \\ 
 0 &   5(2)  A+  &   6(1)  A+   &  201445.6  &  3.1 &     &  \bf{6.0} &      &      &      &  5.4 &      \\ 
 0 &   1(1)  E   &   0(0)  E    &  213427.1  &  3.2 &     &   5.0 &      &      &      &  3.6 &      \\ 
 0 &   5(1)  E   &   4(2)  E    &  216945.6  &  4.1 &  3.3&   3.1 &      &      &  3.7 &  3.4 &      \\ 
 0 &  10(-3)  E  &  11(-2)  E   &  232945.8  &  3.2 &     &   3.7 &      &      &      &  4.5 &      \\ 
 0 &   4(2)  A-- &   5(1)  A--  &  234683.5  &      &     &   5.1 &      &      &      &  5.0 &      \\ 
 0 &   4(2)  A+  &   5(1)  A+   &  247228.7  &  3.1 &     &  \bf{6.2} &      &      &      &  5.6 &      \\ 
 1 &   5(1)  A+  &   6(2)  A+   &  263793.9  &      &     &   4.1 &      &      &      &  3.8 &      \\ 
 1 &   5(1)  A-- &   6(2)  A--  &  265224.4  &      &     &   4.1 &      &      &      &  3.5 &      \\ 
 0 &   6(1)  E   &   5(2)  E    &  265289.6  &  4.1 &  3.3&       &      &      &  3.3 &      &      \\ 
 0 &   9(-3)  E  &  10(-2)  E   &  281956.5  &  3.2 &     &   4.2 &      &      &      &  4.4 &      \\ 
 0 &   3(2)  A+  &   4(1)  A+   &  293464.2  &      &     &   4.7 &      &      &      &  4.5 &      \\ 
 1 &   4(1)  A+  &   5(2)  A+   &  312247.3  &      &     &   4.5 &      &      &      &  3.8 &      \\ 
 1 &   4(1)  A-- &   5(2)  A--  &  313203.4  &      &     &   4.5 &      &      &      &  3.5 &      \\ 
 0 &   8(-3)  E  &   9(-2)  E   &  330794.0  &  3.1 &     &   4.5 &      &      &      &  4.1 &      \\ 
 1 &   3(1)  A+  &   4(2)  A+   &  360661.6  &      &     &   4.8 &      &      &      &  3.8 &      \\ 
 1 &   3(1)  A-- &   4(2)  A--  &  361236.5  &      &     &   4.7 &      &      &      &  3.5 &      \\ 
 0 &   7(-3)  E  &   8(-2)  E   &  379494.1  &      &     &   4.7 &      &      &      &  3.6 &      \\ 
 1 &   2(1)  A+  &   3(2)  A+   &  409035.5  &      &     &   4.8 &      &      &      &  3.9 &      \\ 
 1 &   2(1)  A-- &   3(2)  A--  &  409323.4  &      &     &   4.7 &      &      &      &  3.4 &      \\ 
 0 &   6(-3)  E  &   7(-2)  E   &  428087.7  &      &     &   4.5 &      &      &      &* 3.1 &      \\ 
 1 &   1(1)  A+  &   2(2)  A+   &  457367.8  &      &     &   4.9 &      &      &      &  4.1 &      \\ 
 1 &   1(1)  A-- &   2(2)  A--  &  457463.9  &      &     &   4.7 &      &      &      &  3.6 &      \\ 
 0 &   5(-3)  E  &   6(-2)  E   &  476600.9  &      &     &   4.0 &      &      &      &      &      \\ 
 1 &   4(-2)  E  &   5(-3)  E   &  633572.0  &      &     &   4.1 &      &      &      &  3.2 &      \\ 
 1 &   3(-2)  E  &   4(-3)  E   &  681789.7  &      &     &   4.4 &      &      &      &  3.3 &      \\ 
 1 &   2(-2)  E  &   3(-3)  E   &  730013.6  &      &     &   4.5 &      &      &      &  3.4 &      \\ 
\hline\\                          
\end{tabular}}                          

\end{table*}


Sobolev \etal\ (1997b, hereafter SCG97b) examined the model predictions under a range of circumstances, and presented a list of class~II methanol maser candidates.  These included all the class~II maser transitions known at the time, a few which have been identified in subsequent observations (e.g. 85.568, 86.615 and 86.902~GHz by Cragg \etal\ 2001; 108.893~GHz by Val'tts \etal\ 1999), some lines which have been observed in maser sources but which have not been found to display maser characteristics, and many further lines which have yet to be investigated.  Table~\ref{tab:predictions} presents an updated list, based on calculations with the new collision rate coefficients of PFD.  The range of model conditions chosen differs somewhat from those presented in SCG97b following our experience in fitting models to observations.  

Table~\ref{tab:predictions} tabulates the masers for 8 selected models.  All models have dust temperature $\Td>100$~K and specific column density of methanol $11.5 \leq \scdm \leq 12.0$ in the range which generate significant class~II methanol maser intensities.  They include a range of gas temperatures and densities to illustrate the effects of the collision model.  Although the majority of 6.6-GHz maser sites have no associated uc\HII\ continuum, there is evidence that the weaker maser transitions are more readily detectable in sources in which the uc\HII\ region has already developed (Ellingsen \etal\ 2004), and so all but one of the selected models include background continuum radiation.  The list contains all transitions which attain brightness temperature above $10^{4}$~K in any of the selected models.  Some masers require rather specific conditions to become active, while others are bright across the full range of models illustrated.  To make these trends more apparent, brightness temperatures above $10^{6}$~K are shown in bold, while blank entries mean $\Tb<10^{3}$~K.  The 6.6- and 12.1-GHz masers are usually the strongest, both in observations and the models, and we refer to the other transitions as the weaker class~II masers.  The detectability of these weaker masers is governed by their (unknown) size, which determines the flux density; e.g. Slysh \etal\ (1999) obtained a lower limit of $5 \times 10^{5}$~K for the brightness temperature of the strongest 107.0-GHz maser spots in W3(OH).

Model~1 is an example with low gas density where the masers are influenced by radiative effects alone.  Since the gas temperature is irrelevant in these circumstances, it is equivalent to Model~7 of SCG97b.  The same masers are present in the list as previously, although their brightness is reduced slghtly, which may be attributed to the inclusion of more energy levels in the current modelling (see Fig.~\ref{fig:vary_vt}).  Model~2 represents warm gas (\Tk=150~K) of moderately high density ($\nH=10^{7}$~\cc), where the collisions have begun to quench out the masers (e.g. there are only 3 transitions with $\Tb>10^{6}$~K, compared with 13 transitions in Model~1).  Model~3 illustrates the cool gas (\Tk=30~K), warm dust (\Td=175~K) combination which generates the largest number of intense masers (24 have $\Tb>10^{6}$~K).  These conditions are identical with Model~6 of SCG97b.  A detailed comparison shows that the same masers are present in both cases, some being a little stronger with the improved modelling while others are a little weaker, due primarily to the effects of the new collisions.  The maser candidates listed for Models 1--3 are expected to be most readily detected in sources with masers at 107.0~GHz, which is the most prevalent of the weaker masers.

Models~4 and 5 illustrate conditions which may be more typical of class~II maser sources, since they meet the observed upper limits on 23.1- and 107.0-GHz masers from Cragg \etal\ (2004) (see Section 3.3).  Model~4 has no background uc\HII\ continuum while Model~5 includes it; the effect is to enhance the maser brightness temperatures, particularly at the lower end of the frequency range.  The list of strong masers with $\Tb>10^{6}$~K is confined to the widely distributed 6.6- and 12.1-GHz transitions, plus the (difficult to observe) 57.032- and 60.531-GHz transitions.  

Model~6 is obtained from fitting millimetre wavelength maser observations in W3(OH), and is the same as Model~E from Table~\ref{tab:w3oh}.  Again it illustrates warm gas at moderately high density, and since in this case the gas and dust temperatures are equal, most of the masers present do not develop very strong brightness temperatures.  Model~7 illustrates a combination of masers similar to those observed in NGC~6334F, including intense masers at 19.9 and 23.1~GHz, generated by a warm dust/cool gas combination.  

Finally, Model~8 has dust temperature closer to the threshold for class~II maser activity, and exhibits weak maser action in several transitions absent from the other models.  These include the 25-GHz J$_{2}$-J$_{1}$~E class~I maser series, and transitions at 9.936, 20.171, 23.444 and 104.300~GHz which develop prominently in  models with temperatures in between the class~II and class~I methanol maser domains.  The transitions probably characterize a new \lq intermediate\rq\ class of methanol masers, perhaps exemplified in W33-Met (Voronkov \etal\ 2004).  Of these lines tracing intermediate-class masers, only the 9.9-GHz line has been searched for, and only in a very limited number of sources.  Slysh, Kalenskii \& Val'tts (1993) detected a weak 9.9-GHz maser in W33-Met, but obtained no detections from a search of 5 class~II maser sources. 


\section{DISCUSSION}

A good example of the distortions which can arise from model calculations including too few energy levels is presented by Sobolev \& Deguchi (1994).  In their Model D1, which includes levels up to J=12 for torsionally excited states \vt=0,1,2, the 12.1-GHz \twelve\ transition becomes a strong maser with $\Tb=4.7 \times 10^{10}$~K under warm dust/cool gas conditions.  However, Model D1a, which employs the same physical parameters but omits the \vt=2 levels, instead produces absorption in the 12.1-GHz line, with $\Tb=-1.1 \times 10^{4}$~K.  Note that there is considerable overlap in energy between the \vt=1 and \vt=2 levels (Fig.~\ref{fig:levels}).  Thus the switch-on of the class~II masers at dust temperatures above 100~K is clearly attributable to excitation to the levels of the second torsionally excited state.  In contrast, we have shown that further extending the energy level set to include levels of \vt=3, \vt=4 and the CO-stretch vibration (\vCO=1) has a minor effect by comparison.  We find the class~II masers to be adequately modelled for dust temperatures up to 300~K when levels up to J=18 and \vt=3 are included.  At higher dust temperatures excitation to the \vCO=1 levels tends to quench the masers, and it is likely that the levels of the CH$_{3}$ rock vibration will also play some role.  The indications from infrared observations are that dust temperatures in class~II methanol maser souces are in the range 100--200~K (De Buizer \etal\ 2000), consistent with these findings. 

The accurately calculated collision rate coefficients of PFD provide a much firmer footing for excitation models of both maser and quasi-thermal methanol.  It will be particularly important to reassess the models of class~I methanol masers in the light of the new data, since collisions are thought to provide the source of energy for excitation.  Nevertheless, for both the class~I and class~II masers, it is the radiative transitions which are decisive for generating the population inversions (Cragg \etal\ 1992).  Thus it is not surprising that we find the list of transitions which become class~II masers to be hardly altered by the new collisions.  What has changed is the details of conditions under which individual masers are thermally quenched, and in particular the upper limit on gas density in the 6.6-GHz maser regime, which has increased by approximately a factor of 10 to $10^{9}$\cc.  As before, the observational limits suggest the majority of 6.6-GHz sources (those without accompanying masers at 23.1 or 107.0~GHz) are either near the high density limit of maser range (now $10^{7}-10^{9}$~\cc), and/or near the low dust temperature threshold ($100<\Td<150$~K).  So our conclusions regarding the majority of class~II methanol maser sources are qualitatively unchanged.

The updated list of maser candidates (Table~\ref{tab:predictions}) includes 108 transitions, of which 83 are in common with the corresponding table of SCG97b.  The 14 transitions in the list of SCG97b which are absent from Table~\ref{tab:predictions} are mostly close to the threshold for inclusion ($\Tb=10^{4}$~K).  Similarly, a few of the additional candidates in this new list are rather close to this threshold.  The remainder of the new candidates are included by virtue of their appearance under the conditions of Model~8, suggestive of an intermediate regime spanning the switch between class~I and class~II, which was not examined in SCG97b.  While the collision model may be further improved in the future if rate coefficients for higher J values and for ortho-H$_{2}$ become available, it is encouraging that the potentially observable maser lines can be identified irrespective of the knowledge of accurate collisional data.

It is readily apparent from Table~\ref{tab:predictions} that the warm dust/cool gas combination (Models 3 and 7) has the greatest potential for producing class~II masers.  Since this combination also produces very intense masers at 6.6 and 12.1~GHz, the original modelling of Sobolev \& Deguchi (1994) focussed on this regime.  However, subsequent surveys at several candidate frequencies have shown the weaker maser transitions to be rarely detectable.  There are 25 sources with masers detected in the $3_{1}-4_{0}\ \rm{A}^{+}$ transition at 107.0~GHz out of more than 175 class~II methanol maser sources surveyed (Val'tts \etal\ 1995, 1999; Caswell \etal\ 2000; Minier \& Booth 2002), as well as a handful of exceptional sources with masers known also at several other frequencies.  While the conditions in these sources are clearly not representative of the majority, they provide the opportunity for multitransition fitting to constrain those conditions, as illustrated earlier for W3(OH).  Here the availability of interferometric data makes the results more reliable.  Without this we cannot be confident that the masers at different frequencies are coincident, as is assumed by the modelling.  

The models of the northern maser component in W3(OH) are consistent with a warm dust/warm gas regime, confirmed above using the new colliisons.  The 108.8-GHz $0_{0}-1_{-1}\ \rm{E}$ transition is significantly affected by the improved PFD collision model, not surprisingly since it comes from levels so low in energy.  We have investigated trends in models with parameters in the same vicinity as those derived previously, without fully accounting for the observed upper limit on this maser.  Further exploration of the models may yield a better fit, but some of these require new physical assumptions which are beyond the scope of this paper (e.g. different dust models as in Ostrovskii \& Sobolev 2002).  Nevertheless, the detection of strong masers at 19.9, 23.1, 37.7 and 85.5~GHz in a few sources is evidence of rather different conditions elsewhere, for example in NGC~6334F, G345.01+1.79, and the southern source in W3(OH).  With the new collision data there may be new possibilities to explain these observations, and we will consider this in more detail in future work together with additional observational data providing better constraints.


\section{CONCLUSION}

Since the reliability of the model calculations for class~II methanol masers is influenced by the completeness and accuracy of the fundamental molecular data employed, we have undertaken new model calculations to ascertain firstly the effects of including more energy levels, and secondly the effects of the new collision rate coefficients.  The improved molecular data does not lead to dramatic changes in the models, as the maser pumping is dominated by radiative effects, and the previous models included the essential processes.  The list of approximately 100 transitions which are potential class~II methanol maser candidates is  litle changed and may be regarded as firmly established.  Energy levels up to \vt=3 and J=18 are found to be sufficient for models with dust temperatures up to 300~K.  The new collisional rate coefficients of PFD do affect the absolute and relative brightnesses of the masers over the intermediate density range $10^{4}-10^{9}$~\cc.  But at gas temperatures above 100~K the changes are broadly equivalent to a reduction of the collision cross-section, which may be compensated by an increase in the gas density.  Thus previous conclusions regarding the nature of the most common 6.6-GHz class~II maser sources should be modified to accommodate a slightly higher range of gas densities (maximum $10^{9}$~\cc).  Fits of the model to observations in special sources exhibiting maser emission at many frequencies are modified in detail by the new data, but the example investigated shows that a comparable fit can be obtained under nearby conditions to those previously identified.  The newly available data therefore allows the modelling to proceed with increased confidence.


\section*{Acknowledgements}

Thanks to David Flower for useful discussions of how best to employ the new collision rate coefficients for methanol.  DMC and PDG acknowledge financial support for this work from the Australian Research Council.  AMS was supported by RFBR grant no. 03-02-16433.


\appendix

\section{Methanol energy levels and transitions}

Fig.~\ref{fig:levels} displays energies of E-species methanol up to approximately 600~K in the ground torsional state \vt=0, together with the corresponding levels for \vt=1,2,3,4, and for the \vt=0 mode of the CO-stretch vibration \vCO=1.  In each torsional or vibrational state 284 rotational levels are included, up to an energy threshold corresponding to maximum J=22 and maximum K=9.  The same number of levels are required for the A species.  The levels are arranged as ladders of increasing J for each value of K.  It is evident that the ladders are spaced in an irregular fashion in the torsionally excited states, with considerable overlap among the different torsional states.  For example, there is no clear demarcation between \vt=1 and \vt=2.  The disorder is a result of the large torsional-rotational interaction in methanol, which also produces irregular spacings in the torsional ground state levels (not apparent on this scale).  In contrast, the \vCO=1 levels more closely resemble the \vt=0 levels.  The \vt=0,1,2,3,4 energies come from Mekhtiev, Godfrey \& Hougen (1999), calculated using the Hamiltonian of Xu \& Hougen (1995), which gives microwave accuracy for the rotational states of \vt=0,1,2.  The energies of the \vCO=1 levels are calculated from the data of Moruzzi \etal\ (1989), using Taylor series fits to observed infrared frequencies, relative to the vibrational ground state energies from Moruzzi \etal\ (1992).  The \vCO=1 levels overlap significantly in energy with both \vt=3 and \vt=4.  The \vt=1 mode of the \vCO=1 vibration (not shown) also overlaps in energy with \vt=4, but data is not available for the complete range of K required for our calculations.  The extensive study of Moruzzi \etal\ (1989) also includes data on the CH$_{3}$ rock levels for a single K value in the A symmetry species: these are at comparable energy to the CO stretch.  We may therefore expect that even the large number of energy levels illustrated (1704 for each symmetry species) will be insufficient under circumstances where the \vCO=1 levels have a significant effect on the maser pumping, because levels of comparable energy are omitted.

The line strengths have been accurately calculated for transitions between rotational levels up to J=22 by Mekhtiev \etal\ (1999) for torsional states \vt=0,1,2.  This includes both transitions within each torsional state, and transitions between the different states.  We have used approximate methods to evaluate the line strengths for the higher excited states.  These should be sufficient to reveal whether or not the higher states are important contributors to the maser pumping, and therefore whether the considerable effort of accurately calcuating more line strengths is warranted.  Line strengths for $\Delta$K=0,$\pm1$ radiative transitions within \vt=3, and within \vt=4, were approximated by the corresponding symmetric rotor line strength or H\"{o}nl-London factor for transitions between states of the same J and K (e.g. Zare 1988, Table 6.6).  Line strengths for $\Delta$K=$\pm1$ transitions between \vt=3 or \vt=4 and the other torsional states were set to the product of the symmetric rotor line strength with the square of the torsional overlap.  The torsional overlap between states of different \vt\ and K was obtained from the eigenstates of the torsional Hamiltonian of De Lucia \etal\ (1989) (incorporating the Wang transformation for the A-species to obtain the correct  $\pm$ symmetry states).  The Einstein~A coefficients which govern the radiative transition probabilities are calculated from the line strengths, and depend also on the square of the relevant dipole moment component and the cube of the frequency.  Both the inter- and intra-torsional transitions are governed by the permanent dipole moment components: $\mu_{a}=0.896$~Debye for $\Delta$K=0 transitions and $\mu_{b}=1.412$~Debye for $\Delta$K=$\pm1$ transitions (Sastry, Lees \& Van der Linde 1981).  For transitions among the \vCO=1 levels we used the same line strengths as for the corresponding \vt=0 transitions, and the same dipole moment components.  Since the CO stretch takes place (mainly) along the a-axis, the major transitions between \vCO=1 and \vCO=0 will be transitions to \vt=0 with $\Delta$K=0, and these were again approximated by the symmetric rotor line strengths.  For the CO-stretch transition dipole moment we used the value 0.27~Debye obtained experimentally by Henningsen \& Petersen (1988).


\section{Rate coefficients for collisions}

The methanol collision partners are assumed to be 20 percent He and 80 percent para-H$_{2}$.  Pottage \etal\ (2002, 2004b) tabulate rate coefficients for \vt=0 levels up to J=9 at 13 temperatures between 5 and 200~K, for both collision partners and both A and E symmetry species of methanol.  Note that the state designation used by PFD is not exactly the same as used here; for example E0 and E1 refer to E species \vt=0 and \vt=1 levels respectively, with A0 and A1 defined in a corresponding fashion.  For the A species the sign of K used in the most recent tabulations by PFD matches the $\pm$ symmetry label.  We used linear interpolation between adjacent tabulated values for intermediate temperatures, with  scaling by $\sqrt{T}$ for extrapolation to higher temperatures.  When there are near-degenerate levels, as occurs for high J and K in the A species, the data presented by PFD includes rate coefficients for only one state of the degenerate pair; the same rate coefficients were used for the other state.  The rate coefficient for $\Delta$J=0=$\Delta$K asymmetry doublet transitions between near-degenerate levels was set equal to the value for the J=9, K=4 doublet, which is the highest doublet for which PFD distinguish between the levels. 

Since the method of PFD generates less accurate energies than we employ,  the states used in the calculation of rate coefficients must be mapped onto the set of levels used in the excitation calculations.  Some care is required when the states appear in different energy sequence in the two sets, so that an upward transition in one set becomes a downward transition in the other, and vice versa.  Rate coefficients $R$ for upward and downward transitions between any pair of levels i and j obey the detailed balance relation, which ensures that a Boltzmann population distribution at temperature $T$ is recovered in the high density limit.  This may be expressed in terms of the following quantity, which is independent of the sense (up or down) of the transition.

$$g_{\rm{i}} R_{\rm{i \rightarrow j}} \exp(E_{\rm{j}}/k_{\rm{B}}T) = g_{\rm{j}} R_{\rm{j \rightarrow i}} \exp(E_{\rm{i}}/k_{\rm{B}}T)$$ 

where $g$ represents the (2J+1) statistical weight factor, $E$ is the energy, and $k_{\rm{B}}$ is the Boltzmann constant.  This invariant expression was used to rescale both the upward and downward rates, after ensuring that the energies in both sets were referred to the same ground state, in order to account for the different energy values attributed to the states in our formulation, and specifically for differences in the energy sequence.  For the remaining transitions where no data is available from PFD the rate coefficients for transitions downward in energy were defined as described below, with the corresponding upward rate coefficients calculated from detailed balance.

For transitions starting from J$>$9 in \vt=0, the J=9 rate coefficient for the transition from the same initial K to the same final state was used where possible, extrapolated according to the 1/$\mid$$\Delta$J$\mid$ propensity rule.  When both initial and final J are $>$9, the corresponding J=9 rate coefficient for the same $\Delta$J and initial and final K values was used, if necessary scaled by the 1/$\mid$$\Delta$J$\mid$ propensity rule if $\Delta$J exceeds what is available in the PFD data.  Since our calculations include levels up to either J=18 or J=22, there are a large number of collisional transitions which use these \lq extrapolated\rq\ rate coefficients, but it is not so critical to have accurate values as these higher J levels carry only a small portion of the population. 

Transitions within \vt=1 for He interacting with A-species methanol were treated the same as \vt=0.  Data is available for $T=20$~K only, and was extrapolated to other temperatures using the $\sqrt{T}$ rule.  The He data was assumed in this case to represent para-H$_{2}$ also.  The corresponding E-species data was not used as it is insufficiently accurate (D. Flower, private communication). 

Rate coefficients for rotational transitions within all other torsionally excited states, and within \vCO=1, were based on the propensity rules.  These are similar in form to those of Lees \& Haque, but based on a detailed examinination of the PFD data for \vt=1 A species.  At temperature $T=20$~K the rate coefficient $3\times 10^{-11}$~\rate\ was used for $\Delta$J=1, $\Delta$K=0 transitions, together with $1\times 10^{-11}$~\rate\ for $\Delta$J=0, $\mid$$\Delta$K$\mid>$0 transitions.  In both cases 1/$\mid$$\Delta$J$\mid$ scaling was used for larger values of $\Delta$J.  The value $3\times 10^{-11}$~\rate\ was used for the A species $\Delta$J=0=$\Delta$K asymmetry doublet transitions.  Rate coefficients at different values of temperature were scaled by $\sqrt{T}$. 

Pottage \etal\ (2004a) calculated cross-sections for transitions between \vt=1 and \vt=0, but not rate coefficients.  Cross-sections for torsionally inelastic transitions were found to be typically two orders of magnitude smaller than cross-sections for torsionally elastic transitions.  Furthermore they exhibited no clear propensity for particular $\Delta$J or $\Delta$K.  Based on the typical magnitudes of the \vt=0 rate coefficients, we adopted the value $5 \times 10^{-14}$~\rate\ for all downward transitions involving changes in \vt\ or \vCO.   



\end{document}